\documentclass[a4paper,11pt]{article}
\usepackage{pos}
\usepackage{braket}

\newcommand*{\cf}{cf.\ }
\newcommand*{\eg}{e.\,g.\ }
\newcommand*{\ie}{i.\,e.\ }

\title{New multiloop capabilities of FeynCalc 10}
\ShortTitle{New multiloop capabilities of FeynCalc 10}
\author*[a]{Vladyslav Shtabovenko}

\affiliation[a]{Vladyslav Shtabovenko \\
	Center for Particle Physics Siegen (CPPS), Theoretische Physik 1,  \\
	Universität Siegen, Walter-Flex-Str. 3,  \\
	57068 Siegen, Germany}

\emailAdd{shtabovenko@physik.uni-siegen.de}

\abstract{
	We briefly introduce new multiloop capabilities of the \textsc{Mathematica}
	package \textsc{FeynCalc} 10 and a collection of interfaces connecting \textsc{FeynCalc}
	to such popular tools as \textsc{QGRAF}, \textsc{Fiesta}, \textsc{pySecDec}, \textsc{LoopTools}, 
	\textsc{KIRA}, \textsc{FIRE} or \textsc{Fermat}. In addition to that, we showcase the application of these codes to the
	ongoing study of the ``soft-overlap'' contribution to $B_c \to \eta_c$ transition form factors
	at large hadronic recoil.	
}

\FullConference{Loops and Legs in Quantum Field Theory (LL2024)\\
	14-19, April, 2024\\
	Wittenberg, Germany
}

\begin{document}
\maketitle

\section{Introduction}	

Multiloop calculations belong to the fields of research that heavily depend on software tools and the available 
computational resources. Obviously, the majority of steps necessary to calculate a Feynman diagram involves
repetitive algebraic manipulations that are highly amenable to automation. However, the number
of the required manipulations as well as the number of Feynman diagrams at higher perturbative
orders tend to grow so fast, that any attempt to approach such calculations in a naive way will inevitably
break down. This is why the development of public codes implementing state-of-the-art algorithms and streamlining the 
common steps required to perform higher-order perturbative calculations is crucial to enable further progress
in the field.

While automation of tree- and  one-loop-level calculations has been successfully pushed forward since several
decades (\cf \eg refs.~\cite{Alwall:2014hca,Cullen:2011ac,GoSam:2014iqq,Bahr:2008pv,Bellm:2015jjp,Bevilacqua:2011xh,Nason:2004rx,Frixione:2007vw,Alioli:2010xd, Gleisberg:2008ta,Sherpa:2019gpd, Moretti:2001zz,Kilian:2007gr,Belyaev:2012qa,CompHEP:2004qpa,Yuasa:1999rg,Fujimoto:2002sj,Hahn:1998yk}), the multi-loop case still 
remains more challenging. Despite some progress in the last years \cite{Campbell:2022qmc,Abreu:2020xvt,Heinrich:2023til,Borowka:2016agc,Borowka:2016ehy,Pozzorini:2022ohr,Zoller:2022ewt,Canko:2023lvh}, even at two loops a fully generic code for automatic cross-section and decay rate calculations is still out of reach.

On the one hand, we have plethora of publicly available tools for handling different steps of multiloop Feynman diagram evaluation including a powerful symbolic manipulation system \textsc{FORM} \cite{Vermaseren:2000nd,Kuipers:2012rf} capable of handling expressions containing millions of terms. On the other hand, full automation of such calculations in a manner similar to \eg \textsc{MadGraph} or \textsc{FormCalc} is still far from being attainable due to many technical challenges accompanying such endeavors. 

Partial automation, \ie generation of Feynman diagrams, their algebraic simplification and the reduction of loop
integrals to masters, is something that has been done by the practitioners since many years. Unfortunately, 
in most cases no usable codes were made public. Some famous programs such as \textsc{q2e} and \textsc{exp} from Karlsruhe  \cite{Harlander:1998cmq,Seidensticker:1999bb} are not freely downloadable but at least available upon request.

In recent years this situation started to change with many frameworks addressing automatic calculations 
of multiloop amplitudes being made available to the wider public under open source licenses. Some notable examples
are \textsc{Alibrary}\footnote{\url{https://magv.github.io/alibrary/}}, \textsc{tapir} \cite{Gerlach:2022qnc}, \textsc{FeAmGen.jl}~\cite{Wu:2023qbr}, \textsc{HepLib}~\cite{Feng:2021kha,Feng:2023hxy} or 
\textsc{MaRTIn}~\cite{Brod:2024zaz}. This is undoubtedly a very positive development in our field that has potential
to make such calculations more accessible to the vast majority of particle theorists.

In this proceeding we would like to report on another tool that falls into this category, known under the name of \textsc{FeynCalc} \cite{Mertig:1990an,Shtabovenko:2016sxi,Shtabovenko:2020gxv,Shtabovenko:2021hjx,Shtabovenko:2023idz}.
Unlike most other multiloop codes that were written from scratch, this \textsc{Mathematica} package has been known to the community for almost 35 years. Initially developed as a tool for one-loop calculations, it gradually evolved
towards higher loops, culminating in the official release of \textsc{FeynCalc} 10 at the end 2023. In addition to that, we have developed a collection of interfaces connecting \textsc{FeynCalc} to useful multiloop-related  programs such as \textsc{QGRAF} \cite{Nogueira:1991ex}, \textsc{FIRE} \cite{Smirnov:2014hma,Smirnov:2019qkx,Smirnov:2023yhb}, \textsc{KIRA} \cite{Maierhofer:2017gsa,Maierhofer:2018gpa,Maierhofer:2019goc,Klappert:2020nbg,Lange:2021edb}, \textsc{FIESTA} \cite{Smirnov:2015mct,Smirnov:2021rhf}, \textsc{pySecDec} \cite{Borowka:2017idc} or \textsc{FERMAT} \cite{Lewis:Fermat}. This \textsc{FeynHelpers} add-on for \textsc{FeynCalc} has not yet been officially released but is
already publicly available\footnote{\url{https://github.com/FeynCalc/feynhelpers}} and properly documented.

This report is organized as follows. In Section~\ref{sec:fc} we describe the implementation of \textsc{FeynCalc}'s new multiloop capabilities and briefly mention the related routines. The \textsc{FeynHelpers} add-on is covered in Section~\ref{sec:fh}, while Section~\ref{sec:scet} showcases a practical application of this technology in the context of Soft-Collinear-Effective-Theory (SCET) \cite{Bauer:2000yr,Bauer:2001yt,Beneke:2002ph,Beneke:2002ni}. Finally, in Section~\ref{sec:summary} we summarize the current state of affairs.

\section{FeynCalc} \label{sec:fc}

The most crucial step in making \textsc{FeynCalc} useful for multiloop calculations was to introduce some
sort of topology minimization mechanism. As a typical Feynman amplitude may contain thousands of seemingly
different topologies, finding mappings between those and thus reducing the number of integral families
that need to be IBP-reduced is indispensable.

Here we opted for the so-called Pak algorithm \cite{Pak:2011xt}, an approach that consists of finding one-to-one
mappings between topologies by comparing a particular combination of their Symanzik polynomials $\mathcal{U}$
and $\mathcal{F}$. Normally, when switching to the Feynman parametric representation from the propagator 
representation, the invariance of the integral under loop momentum shifts is translated into the
invariance under renamings of the Feynman parameters $x_i$. Pak algorithm introduces a procedure to determine
a unique ordering of $x_i$ for the given characteristic polynomial $\mathcal{P} = f(\mathcal{U},\mathcal{F})$.
Thus, integral families or single loop integrals can be conveniently compared with each other by calculating
their $\mathcal{P}$ and ordering it according to Pak.

To that aim we introduced the routine \texttt{FCFeynmanPrepare} that determines the Symanzik polynomials
of the given integral or topology. As far as the symbolic representation of the latter is concerned,
loop integrals are called \texttt{GLI}, while topologies are represented using \texttt{FCTopology}
containers. These three building blocks constitute the essence of \textsc{FeynCalc}'s multiloop
functionality.

The two main high level functions on top of that are called \texttt{FCLoopFindTopologyMappings} and 
\texttt{FCLoopFindIntegralMappings}. Given an input in form of \texttt{FCTopology}s or \texttt{GLI}s
they can automatically work out all one-one-to-one mappings detectable by means of the Pak algorithm.
In addition to that, there are many further routines for manipulating input expressions containing
\texttt{GLI} and/or \texttt{FCTopology} symbols. For further information we refer to the official manual 
available as a PDF file \footnote{\url{https://github.com/FeynCalc/feyncalc-manual/releases/tag/dev-manual}}.

In general, given some multiloop amplitude $i\mathcal{M}$ the stages of calculating it with 
\textsc{FeynCalc} will look as follows

\begin{enumerate}
	\item Simplify $i \mathcal{M}$ using \texttt{DiracSimpliy}, \texttt{SUNSimplify} etc.
	\item Identify the occurring topologies with \texttt{FCLoopFindTopologies}
	\begin{itemize}	
		\item In the case of an overdetermined set of propagators use \texttt{FCLoopCreatePartial\-Fractioning\-Rules}
		\item If the set of propagators is incomplete, employ \texttt{FCLoopBasisFindCompletion}
	\end{itemize}	
	\item To map smaller topologies onto bigger ones first find all nonvanishing subtopologies via \texttt{FCLoop\-Find\-Subtopologies}
	
	\item Minimize the number of the topologies using \texttt{FCLoopFindTopologyMappings}
	\item Apply the mappings to $i \mathcal{M}$ and eliminate all integrals in favor of \texttt{GLI}s using \texttt{FCLoop\-Apply\-Topology\-Mappings}
	\item Do the IBP reduction for the occurring \texttt{GLI}s using external tools (\eg \textsc{FIRE} or \textsc{KIRA})
	\item Insert the reduction tables into $i \mathcal{M}$
	\item Check for one-to-one mappings between master integrals with \texttt{FCLoop\-Find\-Integral\-Mappings}
	\item Insert analytic or numerical results for the master integrals
\end{enumerate}

Although the complexity of a typical multiloop calculations might be too high for the capabilities
of \textsc{Mathematica}, in some cases \textsc{FeynCalc} and \textsc{FeynHelpers} alone could be still
sufficient to obtain the final result within a reasonable time frame. Moreover, the multiloop related
routines can be of course also used in other contexts, e.g. when doing most of the calculation in
\textsc{FORM} but still employing \textsc{FeynCalc} at some intermediate steps.

Many ideas behind this implementation were adopted from the thesis of Jens Hoff~\cite{Hoff:2015kub} that
contains a very detailed description of Pak's algorithm and related ideas. Also his unfinished
\textsc{Mathematica} package \texttt{TopoID}\footnote{\url{https://github.com/thejensemann/TopoID}} was enormously
useful for our purposes. The main algorithm of \textsc{FCFeynmanPrepare} was taken from
\textsc{FIRE}'s \texttt{FindRules} routine.

\section{FeynHelpers} \label{sec:fh}

The main motivation behind the development of \textsc{FeynHelpers}~\cite{Shtabovenko:2016whf} was the following observation.
Even though the multiloop community is very prolific in terms of software tools for automatizing 
different aspects of higher order perturbative calculations, the practical applications of such
codes in real-life projects is not entirely straightforward. One of the reasons is that
formats of configuration files as well as input and output expressions vary from tool to tool.
Using the output of one code as an input for another code always requires some conversion steps
that are tedious to implement and require a formidable amount of \textit{glue scripts} written
in \textsc{bash}, \textsc{Python}, \textsc{Mathematica} or other suitable languages.
Owing to the popularity of \textsc{FeynCalc} among particle physics practitioners we deemed that
the format used in this package (in particular with the new \texttt{GLI} and \texttt{FCTopology} symbols)
can be used as a common denominator for exchanging results between various programs.

\textsc{FeynHelpers} is implemented as a collection of interfaces between \textsc{FeynCalc} and some
selected tools that are commonly used in multiloop calculations. As of now we support \textsc{QGRAF}, \textsc{Package-X} \cite{Patel:2015tea,Patel:2016fam}, \textsc{LoopTools} \cite{Hahn:1998yk},
\textsc{FIRE}, \textsc{KIRA}, \textsc{FIESTA}, \textsc{pySecDec} and \textsc{FERMAT}. The add-on provides high-level functions that accept input in \textsc{FeynCalc} format and 
allow processing it either by directly calling the corresponding tool in the background or 
by generating scripts for running that tool either locally or on a cluster. Options can be 
used to steer the evaluation process or to adjust tool's settings. For further information we 
refer to the official manual available as a PDF file \footnote{\url{https://github.com/FeynCalc/feynhelpers-manual/releases/tag/dev-manual}}.

To be more specific here, let us explain how \textsc{FeynHelpers} can be used to perform
an IBP-reduction of loop integrals that were obtained in some calculation and then loaded
into \textsc{FeynCalc} as a list of \texttt{GLI}s with the corresponding \textsc{FCTopology}s.

In the first step we need to supply the topologies to the function
\texttt{FIREPrepareStartFile}. This routine will generate
\textsc{Mathematica} scripts for analyzing the given topologies using \textsc{LiteRed}~\cite{Lee:2013mka}
 (one script per topology) and for creating \texttt{.start} or \texttt{.sbases} and \texttt{.lbases}
files needed for the reduction. Depending on the complexity of the topology  running those scripts can 
take a significant amount of time and should be in general done on a cluster. However, for simple
cases where the whole process only takes few seconds, it is also possible to run them directly from 
an evaluation notebook by calling \texttt{FIRECreateStartFile}.

Then, we also need to create the \texttt{.config} file for the reduction as well as the list of integrals
being reduced. This can be handled via \texttt{FIRECreateConfigFile} and \texttt{FIRECreateIntegralFile} respectively.
The reduction itself should be definitely done on a cluster, but again as a matter of convenience for
very simple cases we also offer a function \texttt{FIRERunReduction} that will start it as a background
process directly  from a \textsc{Mathematica} notebook. Finally, using \texttt{FIREImportResults} we can load the reduction tables into our notebook and convert them into a list of replacements rules where \texttt{GLI}s are substituted by a linear combination of simpler master integrals (also in the \texttt{GLI} format).

We would like to stress that this interface can be useful also in \textsc{FORM}-based setups where one 
needs to prepare \textsc{FIRE} runcards for a large number of integral families obtained in the course
of some calculation completely unrelated to \textsc{FeynCalc}. The only thing one needs to do is to convert
the occurring integral families and loop integrals into the \texttt{FCTopology} and \texttt{GLI} formats 
respectively.

\section{Structure of soft-overlap contribution to \texorpdfstring{$B_c \to \eta_c$}{bc to etac} form factors} \label{sec:scet}

The presented tools have already been employed in a real-life multiloop calculation, where
we were interested in obtaining a better understanding of QCD factorization with a systematic 
inclusion of power corrections. Although power corrections can be studied in the framework of SCET, some effects appearing at subleading power, in particular the end-point divergent convolution integrals, still remain problematic. As has been shown recently \cite{Bell:2022ott}, the possible remedy in form of refactorization \cite{Boer:2018mgl,Liu:2019oav} does not solve all problems in hard-exclusive
processes. To illustrate this point more explicitly, it is useful to consider the $B_c \to \eta_c$ form factors in the nonrelativistic approximation ($m_b \gg m_c \gg \Lambda_{\textrm{QCD}}$). This process can serve as a perfect laboratory to study 
the all-order structure of the associated double-log corrections. In particular, the all-order double-log structure at large recoil can be predicted from solving 
coupled integral equations and then explicitly verified using a method-of-regions analysis 
\cite{Boer:2023tcs}.

\begin{figure}
				\centering
	\includegraphics[width=0.5 \textwidth]{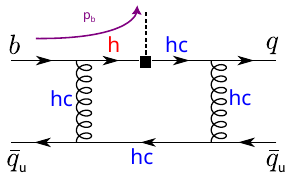}
	\caption{A representative one-loop diagram for the $B_c \to \eta_c$ transition at partonic level. The dashed line represents an emitted $W$-boson, while ``h'' and ``hc'' mean, that the corresponding propagators are hard or hard-collinear respectively. The purple line shows the correct routing of the $b$-quark momentum that respects the assigned power-counting rules. Dressing this diagram with more gluons generates higher order QCD corrections to this process.}
	\label{fig:scet}
\end{figure}

However, in order to check this conjecture at fixed order up to three-loops one extra ingredient
is needed: the purely hard-collinear coefficient $F_{\textrm{hc}}(\gamma)$, with 
\begin{equation}
	F(\gamma) \equiv \frac{1}{2 E_\eta} \braket{\eta_c (p_\eta) | \bar{c} \Gamma b| B_c (p_B) }.
\end{equation}
The building block $F_{\textrm{hc}}(\gamma)$ has to be explicitly extracted from the corresponding diagrams 
evaluated at two and three loops, where multiloop techniques become indispensable. To be more specific,
we need to consider two and three-loop QCD corrections to the process (\cf Figure~\ref{fig:scet})
\begin{equation}
	b ( m_b v^\mu ) \, \bar{q}_u (m_{\bar{q}_u} v^\mu ) \to W(q_1) \, q ( m_{q} {v'}^{\mu} ) \, \bar{q}_u (  m_{q} {v'}^{\mu}),
\end{equation}
where $\bar{q}_u$ denotes an up-type antiquark, while $q$ stands for some other light quark. In a physical $\eta_c$ meson
one obviously has $\bar{q}_u = \bar{c}$ and $q =c$, but here for pedagogical purposes we choose to treat them as different species. The kinematics is chosen such that
\begin{align}
	v^\mu &= \frac{n^\mu + \bar{n}^\mu}{2}, \quad v^2 = 1, \\
	{v'}^{\mu} &=\gamma n^\mu + \frac{\bar{n}^\mu}{4 \gamma}, \quad (v')^2 = 1, \\
	\gamma & \equiv v  \cdot {v'}, 
\end{align}
where $n$ and $\bar{n}$ are two light-like reference vectors that satisfy
\begin{align}
\quad n^2 = \bar{n}^2 = 0, \quad n \cdot \bar{n} =2,
\end{align}
so that every four-vector $k$ can be decomposed into
\begin{align}
	k^\mu = \frac{\bar{n}^\mu}{2} (k \cdot n)  + \frac{n^\mu}{2} (k \cdot \bar{n}) + k_\perp^\mu \equiv k^\mu_+ + k^\mu_- + k_\perp^\mu.
\end{align}
To extract the limit we are interested in, all loop momenta $k_i$ are considered to be hard-collinear, \ie their
components scale as $(k_+, k_-, k_\perp) \sim (\lambda^2, 1, \lambda)$. The scaling of the remaining quantities is as follows
\begin{align}
	m_\eta \sim \lambda^2, \quad \gamma \sim \frac{1}{\lambda^2}, \quad m_b \sim 1, \quad 
	k_i \cdot n \sim \lambda^2, \quad
	k_i \cdot \bar{n} \sim 1, \quad
	k_i^2 \sim \lambda^2.
\end{align}
To ensure the correct scaling of the $b$-quark propagators, it is crucial, that the hard momentum $p_b =m_b v^\mu$ is always routed through 
the internal $b$-lines and the external $W$-line. Letting it flow through gluon or other light quark propagators would spoil the power counting and lead to wrong results.

To decrease the number of scales entering the calculation it is useful to introduce the dimensionless mass ratios
\begin{align}
	\bar{u}_0 \equiv \frac{m_{\bar{q}_u}}{m_\eta}, \quad u_0 \equiv \frac{m_q}{m_\eta} = 1 - \bar{u}_0,
\end{align}
where a physical $\eta_c$ would have
\begin{align}
	u_0 = \bar{u}_0 = 1/2.
\end{align}
This way we can eliminate the light quark masses using
\begin{align}
	m_q = (1 -\bar{u}_0) m_\eta, \quad m_{\bar{q}_u} = \bar{u}_0 m_\eta,
\end{align}
so that the final result will depend only on $\bar{u}_0$, $m_\eta$, $m_b$ and $\gamma$. Luckily, for our purposes we only need the coefficient in front of the leading pole of the amplitude,
which reduces the number of scales even further. Effectively, only the $\bar{u}_0$-dependence of master integrals is nontrivial at leading power.

We perform this calculation using an automatized setup that employs \textsc{FORM} and \textsc{FeynCalc}. The code itself
is called \textsc{LoopScalla} and will be published in near future. A preliminary version thereof is already available
online\footnote{\url{https://github.com/FeynCalc/LoopScalla}}. In this setup we use \textsc{QGRAF} to generate the required diagrams and \textsc{FORM} for the insertion of Feynman rules. Dirac and color algebra simplifications as well as the expansion of the amplitudes in the hard-collinear limit are also done using \textsc{FORM}. Having extracted all naive topologies appearing in the amplitude we switch to \textsc{FeynCalc} for the purpose of finding mappings between different integral families, performing partial fractioning for cases with overdetermined propagator bases and adding extra propagators when a basis is incomplete. In addition to that, 
\textsc{FeynCalc} also generates rules for rewriting loop momentum-dependent scalar products in terms of inverse propagator denominators. 

The results of this calculational step are exported as \textsc{FORM} \texttt{id}-statements and the insertion of the topology
mappings into preliminary results is done in \textsc{FORM}. Then, after having extracted the final list of loop integrals
for every integral family we use \textsc{FeynHelpers} to generate run cards for \textsc{FIRE}. Upon performing the IBP
reduction we again employ \textsc{FeynHelpers} to load the reduction tables and to export them as \textsc{FORM} \texttt{id}-statements. These reduction rules are then converted into \textsc{FORM} tablebases and finally inserted into 
the amplitudes. 

Then, all master integrals are evaluated numerically using \textsc{pySecDec} in order to determine the leading power of the $\varepsilon$-pole in each of them. Substituting those results into the final expression significantly decreases the number of
integrals that need to be calculated numerically, since many masters do not contribute to the final result. Instead of calculating the remaining masters analytically, we choose a different semi-numerical approach that exploits the fact that we only need their leading poles\footnote{At three loops also subleading poles of some master integrals can enter the $1/\varepsilon^6$-piece of the full amplitude. However, at two loops the presented approach was fully sufficient.}. Making a rational function ansatz
\begin{align}
	\sum_{i=-|a|}^b c_i \bar{u}_0^i + \frac{1}{1-\bar{u}_0} \sum_{i=-|a'|}^{b'} c'_i \bar{u}_0^i + \frac{1}{(1-\bar{u}_0)^2} \sum_{i=-|a''|}^{b''} c''_i \bar{u}_0^i, \quad a,b,a',b',a'',b'' = 3,4,5
\end{align}
for each leading pole coefficient we evaluate each integral numerically at 22 special points
\begin{align}
\bar{u}_0 = 1/2,1/3,1/4, \ldots 1/9, 2/3, 2/5, \ldots 3/4, 3/5, \ldots
\end{align}
and convert the results into rational numbers using \textsc{Mathematica}'s \texttt{Rationalize} function.
Empirically, we found that numerators and denominators containing prime factors larger than 9 lead to ``bad'' points that should be discarded, while the remaining ``good'' points are kept.  For example, while $5/14$ means that we have a ``good'' point, a point generating $113/167$ gets removed. 

Putting these results together, we can generate a system of linear equations for each master integral and successfully determine all $c_i$, $c_i'$ and $c_i''$ coefficients analytically. These results can be easily cross checked by performing more evaluations
using  \textsc{pySecDec} or calculating some of the simpler integrals analytically. At two loops the final result for the leading pole reads
\begin{align}
	i \mathcal{M}^{(2)}  \sim \frac{1}{\varepsilon^4} \left( 
	C_F^2 \frac{15 \bar{u}_0 + 17 }{\bar{u}_0^3} - C_A C_F \frac{\bar{u}_0 + 5}{2\bar{u}_0^3}\right),
\end{align}
which confirms the diagrammatic analysis done in ref.~\cite{Boer:2023tcs}.

The three-loop calculation is currently under way. As compared to the two-loop case, here we have to deal with 20759 diagrams (722 at two-loops) and 6276 integral families (377 at two-loops). The final results are expected to appear this year.

\section{Summary} \label{sec:summary}

In this talk we highlighted  key features of \textsc{FeynCalc} 10 that implements the long-awaited routines 
for semi-automatic multiloop calculations. We also discussed the issue of interfacing different loop-related 
codes with each other and presented our solution in form of an easy-to-use \textsc{FeynCalc} add-on that tackles
this task for some popular programs. Last but not least, we showed a practical application of this tools to a 
SCET-related problem using our \textsc{FORM}-based framework that employs \textsc{FeynCalc} and \textsc{FeynHelpers}.

\section*{Acknowledgments}

The author would like to thank Rolf Mertig and Frederik Orellana for the longstanding collaboration on the \textsc{FeynCalc} project. He is also grateful to Guido Bell, Philipp Böer, Thorsten Feldmann and Dennis Horstmann for the given opportunity to contribute to the study of QCD power corrections at subleading power. 

Most of the calculations described in this work were done using the TTP/ITP high performance cluster at KIT and the OMNI cluster
at the University of Siegen. The author would like to express his sincere gratitude to Manuel Egner, Martin Lang and Oliver Witzel for their help with the usage of the respective clusters.

This research was supported by the Deutsche Forschungsgemeinschaft (DFG, German Research
Foundation) under grant 396021762 -- TRR 257 ``Particle Physics Phenomenology after
the Higgs Discovery''. This report has been assigned preprint numbers SI-HEP-2024-16 and P3H-24-043.

\bibliographystyle{JHEP}
\bibliography{final.bib}

\end{document}